\documentclass[aps,pre,twocolumn,groupedaddress,showpacs]{revtex4-1}

\usepackage{graphicx}
\usepackage{color}
\usepackage{amsmath}

\def\be{\begin{equation}}
\def\en{\end{equation}}
\def\bea{\begin{eqnarray}}
\def\ena{\end{eqnarray}}

\def\ve{\varepsilon}
\def\n{\nabla}

\newcommand{\av}[1]{\langle{#1}\rangle}

\newcommand{\bi}[1]{\mbox{\boldmath$#1$}}

\begin{document}

\title{Bistable director alignments of nematic liquid crystals confined 
in frustrated substrates}

\author{Takeaki Araki$^{1,2}$ and Jumpei Nagura$^1$}

\affiliation{$^1$Department of Physics, Kyoto University, Kyoto 606-8502, Japan\\
$^2$CREST, Japan Science and Technology Agency, Japan}

\date{\today}

\begin{abstract}

We studied in-plane bistable alignments of nematic liquid crystals 
confined by two frustrated surfaces by means of Monte Carlo simulations 
of the Lebwohl-Lasher spin model. 
The surfaces are prepared with orientational checkerboard patterns, 
on which the director field is 
locally anchored to be planar yet orthogonal between the 
neighboring blocks. 
We found the director field in the bulk tends to be aligned along the 
diagonal axes of the checkerboard pattern, 
as reported experimentally  
[J.-H. Kim {\it et al.}, Appl. Phys. Lett. {\bf 78}, 3055 (2001)]. 
The energy barrier between the two stable orientations 
is increased, when the system is brought to the isotropic-nematic 
transition temperature.  
Based on an elastic theory, we found that the bistability 
is attributed to the spatial modulation of the director field 
near the frustrated surfaces. 
As the block size is increased and/or the elastic modulus is reduced, 
the degree of the director inhomogeneity is increased, enlarging the 
energy barrier. 
We also found that the switching rate between the stable states is 
decreased when the block size is comparable to the cell thickness.

\end{abstract}

\pacs{
64.70.mf,
61.30.Hn, 
64.60.De, 
42.79.Kr 
}

\maketitle

\section{Introduction}

Liquid crystals have been utilized in many applications. 
In particular, they are widely used 
in optical devices such as flat panel displays 
\cite{Rasing_book_2004,Chen_book_2011}. 
Because of the softness of the liquid crystal, 
its director field is deformed by 
relatively weak external fields \cite{deGennes_book}. 
To sustain the deformed state, the external field 
has to be constantly applied to the liquid crystal substance. 
In order to reduce power consumption, 
a variety of liquid crystal systems showing 
multistable director configurations or storage effects 
have been developed \cite{Berreman_JAP_1981,
Clark_APL_1980,Davidson_PRE_2002,Cheng_APL_1982,
Boyd_APL_1980,Scheffer_APL_1984,
ParryJones_APL_2003,
Gwag_APL_2007,
Monkade_EPL_1988,
Barberi_JAP_1998,Jagemalm_APL_1998,
Dozov_APL_1997,
Kim_APL_2001,
Kim_Nature_2002,Kim_APL_2003,Yu_APL_2004,
Tsakonas_APL_2007,Araki_NM_2011,Araki_SM_2013,Lohr_SM_2014}. 
In such systems, 
a pulsed external field can induce permanent changes of the 
director configurations. 
Liquid crystals of lower symmetries, 
such as 
cholesteric, ferroelectric and flexoelectric phases, 
are known to show the storage effects 
\cite{Berreman_JAP_1981,Clark_APL_1980,Davidson_PRE_2002}.

A nematic liquid crystal 
in a simple geometry, {\it e.g.} that  
sandwiched 
between two parallel plates with homeotropic anchoring, 
shows a unique stable director configuration if external 
fields are not imposed. 
By introducing elastic frustrations, 
the nematic liquid crystals can have 
different director configurations 
of equal or nearly equal elastic energy 
\cite{Boyd_APL_1980,Scheffer_APL_1984,ParryJones_APL_2003,
Araki_PRL_2006,
Tkalec_Science_2011,
Araki_NM_2011,Araki_SM_2013,
Araki_PRL_2012,Gwag_APL_2007,Lohr_SM_2014}. 
For instance, 
either of horizontal or vertical director orientation 
is possibly formed 
in nematic liquid crystals confined 
between two flat surfaces of uniformly tilted but oppositely directed 
anchoring alignments \cite{Boyd_APL_1980}.  
Also, it was shown that the nematic liquid crystal confined in 
porous media shows a memory effect \cite{Bellini_PRL_2002}. 
The disclination lines of the director field 
can adopt a large number of trajectories running through the 
channels of the porous medium 
\cite{Araki_NM_2011,Araki_SM_2013}. 
The prohibition of spontaneous changes of the defect 
pattern among the possible trajectories leads to the 
memory effect.

Recent evolutions of micro- and nano-technologies 
enable us to tailor substrates of inhomogeneous 
anchoring conditions, the length scale of which 
can be tuned less than the wavelength of visible light. 
With them, many types of structured surfaces for the liquid crystals 
and the resulting 
director alignments have been reported in the past 
few decades 
\cite{Scheffer_APL_1984,Zhang_PRL_2003,Gwag_APL_2007,Lohr_SM_2014,Murray_PRE_2014,Tsui_PRE_2004,Kim_Nanotech_2002}. 
For example, a striped surface, in which the 
homeotropic and planar anchorings appear alternatively, 
was used to control the polar angle of the 
director field in the bulk \cite{Barbero_JPII_1992,Kondrat_EPJE_2003,
Atherton_PRE_2006}. 

Kim {\it et al.} demonstrated 
in-plane bistable alignments 
by using a nano-rubbing technique with an atomic 
force microscope \cite{Kim_APL_2001,Kim_APL_2003}. 
They prepared 
surfaces of orientational checkerboard patterns. 
The director field in contact to the surfaces 
is imposed to be parallel to the 
surface yet orthogonal between the neighboring domains. 
They found that the director field far from the surface tends to be 
aligned along either of the two diagonal axes of the 
checkerboard pattern. 
More complicated patterns are also possible to prepare \cite{Kim_Nature_2002}. 

In this paper, we consider the mechanism of 
the bistable orientations 
of the nematic liquid crystals 
confined in two flat surfaces of the checkerboard 
anchoring patterns. 
We carried out Monte Carlo simulations 
of the Lebwohl-Lasher spin model \cite{Lebwohl_PRA_1972}
and argued their results 
with a coarse-grained elastic theory. 
In particular, the dependences of the stability of the 
director patterns on the temperature, and the domain size 
of the checkerboard patterns are studied. 
Switching dynamics between the stable configurations 
are also considered.

\section{Simulation model}

We carry out lattice-based Monte Carlo simulations 
of nematic liquid crystals confined by two parallel plates 
\cite{Lebwohl_PRA_1972,Chiccoli_PhysicaA_1988,Cleaver_PRA_1991,Mondal_PLA_2003,
Marguta_PRE_2011,Maritan_PRL_1994,Priezjev_PRE_2001,Araki_NM_2011,
Marguta_PRE_2011}. 
The confined space is composed of three-dimensional 
lattice sites ($L\times L \times H$) 
and it is denoted by $\mathcal{B}$. 
Each lattice site $i$ has a unit 
spin vector $\bi{u}_i$ ($|\bi{u}_i|=1$), 
and the spins are mutually interacting with those at the 
adjacent sites. 
At $z=0$ and $z=H+1$, 
we place substrates, 
composed of two-dimensional lattices.
We put unit vectors $\bi{d}_j$ on the site $j$ on $\mathcal{S}$, 
where $\mathcal{S}$ represents 
the ensemble of the substrate lattice sites. 
We employ the following Hamiltonian for $\bi{u}_i$, 
\bea
\mathcal{H}&=&
-\ve
\sum_{\av{i,j} (\in \mathcal{B})}
P_2(\bi{u}_i\cdot\bi{u}_j)\nonumber\\
&&
-\sum_{i\in \mathcal{B}}P_2(\bi{u}_i\cdot \bi{e})
-w\sum_{\av{i,j},i\in \mathcal{B}, j\in \mathcal{S}}
P_2(\bi{u}_i\cdot \bi{d}_j), 
\label{eq:Hamiltonian}
\ena
where $P_2(x)=3(x^2-1/3)/2$ is the second-order Legendre function and
$\sum_{\av{i,j}}$ means 
the summation over the nearest neighbor site pairs. 
We have employed the same Hamiltonian to 
study the nematic liquid crystal confined in 
porous media \cite{Araki_NM_2011}. 

\begin{figure}
\includegraphics[width=0.45\textwidth]{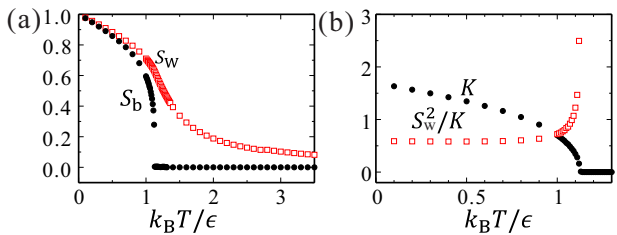}
\caption{
(a) Plots of the scalar nematic order parameter 
in the bulk $S_{\rm b}$ (black circles) 
and on the surface $S_{\rm w}$ (red open squares) with respect to the 
temperature. 
(b) Plot of the elastic modulus $K$ (black circles) 
of the nematic phase with $T$. 
$S_{\rm w}^2/K$ is also plotted with red open squares.
}
\label{fig1}
\end{figure}

The first term of the right hand side of Eq.~(\ref{eq:Hamiltonian}) 
is the Lebwohl-Lasher potential, 
which describes the isotropic-nematic 
transition \cite{Lebwohl_PRA_1972,Chiccoli_PhysicaA_1988,Cleaver_PRA_1991}. 
In Fig.~\ref{fig1}, we plot the temperature dependences of 
(a) the scalar nematic order parameter $S_{\rm b}$ 
and (b) the elastic modulus $K$ 
in a bulk system. 
The numerical schemes for measuring them 
are described in Appendix A. 
We note that 
a cubic lattice with periodic boundary conditions ($L^3=128^3$) 
is used for obtaining $S_{\rm b}$ and $K$ in Fig.~\ref{fig1}.
As the temperature is increased, 
both the scalar order parameter and 
the elastic modulus are decreased and show abrupt 
drops at the transition temperature $T=T_{\rm IN}$, 
which is estimated as 
$k_{\rm B}T_{\rm IN}/\ve\cong 1.12$ \cite{Priezjev_PRE_2001}. 

The second term of Eq.~(\ref{eq:Hamiltonian}) is the coupling between the 
spins in $\mathcal{B}$ and an in-plane external field 
$\bi{e}$. 
The last term represents the 
interactions between the bulk spins and the surface directors, 
that is, the 
Rapini-Papoular type anchoring effect \cite{Rasing_book_2004,deGennes_book}. 
$w$ is the strength of the anchoring interaction. 
If $\bi{d}_j$ is parallel to the substrates and $w>0$, 
the planar anchoring conditions are imposed to the 
spins at the $\mathcal{B}$-sites contacting to $\mathcal{S}$. 
This term not only gives the angle dependence of the anchoring effect 
in the nematic phase, but also enhances the nematic order near the surface. 
In Fig.~\ref{fig1}(a), we also plot the scalar nematic order parameter 
on a homogenuous surface of $w=\epsilon$. 
The definition of $S_{\rm w}$ is described in Appendix A. 
The nematic order on the surface is larger than that in the bulk $S_{\rm b}$
and is decreased continuously with $T$. 
Even at and above $T_{\rm IN}$, 
$S_{\rm w}$ does not vanish to zero. 
When the temperature is far below $T_{\rm IN}$, 
on the other hand, it is close to that in the bulk $S_{\rm b}$.

In this study, we prepare two types of anchoring cells. 
In type I cells, we set hybrid substrates. 
At the bottom surface $(z=0)$, the preferred direction $\bi{d}_j$ 
is heterogeneously patterned like a checkerboard as given 
by 
\bea
\bi{d}_j(x,y)
= \left\{
\begin{array}{ll}
(0,1,0) & 
{\rm if} \;([x/D]+[y/D]) {\rm \; is\; even} \\
(1,0,0) & 
{\rm if} \;([x/D]+[y/D]) {\rm \; is\; odd} \\
\end{array}
\right.,
\label{eq:block}
\ena
where $[X]$ stands for the largest integer smaller 
than a real number $X$.
$D$ is the unit block size of the checkerboard pattern. 
At the top surface $(z=H+1)$, on the other hand, 
the preferred direction is homogeneously set to 
$\bi{d}_j=\bi{d}_{\rm t}\equiv(\cos \phi_{\rm t}\sin \theta_{\rm t},
\sin\phi_{\rm t}\sin\theta_{\rm t},\cos\theta_{\rm t})$. 
$\theta_{\rm t}$ and $\phi_{\rm t}$ are the polar and azimuthal 
angles of the preferred direction at the top surface. 
In type II cells, 
both substrates are patterned like the checkerboard, 
according to Eq.~(\ref{eq:block}). 

We perform Monte Carlo simulations with 
heat bath samplings. 
A trial rotation of the $i$-th spin 
is accepted, considering the local 
configurations of neighboring spins, 
with the probability 
$p(\Delta \mathcal{H})=1/(1+e^{\Delta \mathcal{H}/k_{\rm B}T})$, 
where $\Delta \mathcal{H}$ is the difference of the 
Hamiltonian between before and after the trial rotation. 
The physical meaning of the temporal evolution of Monte 
Carlo simulations is sometimes a matter of debate. 
However, we note that the method is known to be 
very powerful and useful for studying glassy 
systems with slow relaxations, 
such as a spin glass \cite{Ogielski_PRB_1985,Araki_NM_2011}, 
the dynamics of which is dominated by activation processes 
overcoming an energy barrier. 

In this study 
we fix the anchoring strengths at both 
the surfaces to $w=\ve$, for simplicity. 
The lateral system size is $L=512$ and the 
thickness $H$ is changed. 
For the lateral $x$- and $y$- directions, 
the periodic boundary conditions 
are employed.

\section{Results and discussions}

\subsection{Bistable alignments}

\begin{figure}[htbp]
\includegraphics[width=0.45\textwidth]{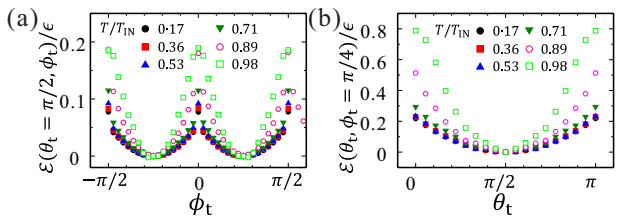}
\caption{
Dependences of the stored energy per unit area 
with respect to 
$\phi_{\rm t}$ at $\theta_{\rm t}=\pi/2$ in (a),  
and to $\theta_{\rm t}$ at $\phi_{\rm t}=\pi/4$ in (b).  
The liquid crystal is confined in a type I cell of $D=8$ and $H=16$. 
The temperature is changed. 
}
\label{fig2}
\end{figure}

First, 
we consider nematic liquid crystals confined 
in cells with the hybrid surfaces (type I). 
Figure~\ref{fig2}(a) 
plots the energies stored in the cell 
with respect to the azimuthal anchoring angle 
$\phi_{\rm t}$. 
Here the polar anchoring angle is fixed to $\theta_{\rm t}=\pi/2$. 
The energy per unit area 
$\mathcal{E}$ is calculated as 
$\mathcal{E}(\theta_{\rm t},\phi_{\rm t})
=\av{\mathcal{H}(\theta_{\rm t},\phi_{\rm t})}/L^2-
\mathcal{E}_{\rm min}$, 
where $\av{X}$ means the spatial average of a variable $X$. 
$\mathcal{E}_{\rm min}$ is the lowest energy defined as 
$\mathcal{E}_{\rm min}=\min_{\theta_{\rm t},\phi_{\rm t}}\av{\mathcal{H}}/L^2$ 
at each temperature (see below).   
$\mathcal{E}$ is obtained after $5\times 10^4$ Monte Carlo steps 
(MCS) in the absence of external fields. 
The cell thickness is $H=16$ and the block size is $D=8$. 
The temperature is changed. 

Figures~\ref{fig2}(a) indicates the 
energy has two minima at $\phi_{\rm t}=\pm \pi /4$, 
while it is maximized at 
$\phi_{\rm t}=0$ and $\pm \pi/2$. 
To see the dependence on 
the polar angle, we plot $\mathcal{E}$ against $\theta_{\rm t}$ 
with fixing $\phi_{\rm t}=\pi/4$ in Fig.~\ref{fig2}(b). 
It is shown that $\mathcal{E}$ 
is minimized at $\theta_{\rm t}=\pi/2$ for $\phi_{\rm t}=\pi/4$. 
Hence we conclude that 
the stored energy is globally lowest 
at $(\theta_{\rm t},\phi_{\rm t})=(\pi/2,\pm \pi/4)$, 
so that we set 
$\mathcal{E}_{\rm min}
=\av{\mathcal{H}(\theta_{\rm t}=\pi/2,\phi_{\rm t}=\pi/4)}/L^2$ 
in Fig.~\ref{fig2}. 
This global minimum indicates that 
the parallel, yet bistable configurations of the director field 
are energetically preferred in this cell. 
This simulated bistability is in accordance with the 
experimental observations reported by Kim {\it et al.} \cite{Kim_APL_2001}. 
When a semi-infinite cell is used, 
the bistable alignments of the director field would be realized. 
Hereafter, we express these two stable directions with 
$\hat{\bi{n}}_+$ and $\hat{\bi{n}}_-$. 
That is, $\hat{\bi{n}}_\pm=(1/\sqrt{2},\pm 1/\sqrt{2},0)$. 

\begin{figure}
\includegraphics[width=0.45\textwidth]{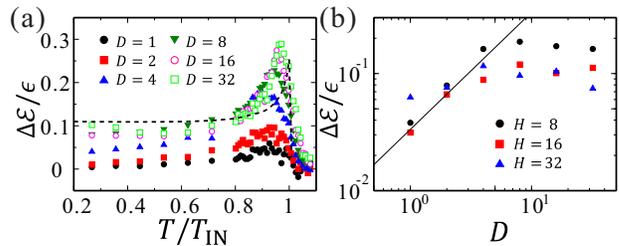}
\caption{
(a)
Plot of the energy difference 
$\Delta \mathcal{E}$ per unit area against the temperature. 
The cell thickness is $H=8$ (type I) and the block size $D$ is changed. 
(b) 
Plot of the energy difference per unit area against the block size. 
The temperature is $T/T_{\rm IN}=0.89$ and the cell thickness 
is changed. }
\label{fig3}
\end{figure}

Figure~\ref{fig2} also indicates the temperature dependencies 
of the stored energies. 
When the temperature 
is much lower than the transition temperature $T_{\rm IN}$, 
the curves of $\mathcal{E}$ are 
are rather flat. 
As the temperature is increased, 
the dependence becomes more remarkable. 
Figure~\ref{fig3}(a) 
plots the energy difference between the 
maximum and minimum of $\mathcal{E}$ 
for fixed $\theta_{\rm t}=\pi/2$ 
as functions of $T$. 
It is defined by 
the in-plane rotation of $\bi{d}_{\rm t}$ 
as $\Delta \mathcal{E}=
\mathcal{E}(\theta_{\rm t}=\pi/2,\phi_{\rm t}=\pi/4)
-\mathcal{E}(\pi/2,0)$.  
We plot them for several block sizes $D$, while 
the cell thickness is fixed to $H=16$. 

In Fig.~\ref{fig3}(a), we observe non-monotonic dependences of 
the energy difference on the temperature. 
$\Delta \mathcal{E}$ is almost independent of $T$ 
when $T/T_{\rm IN}<0.6$. 
In the range of 
$0.6 \lesssim T/T_{\rm IN} < 0.9$, 
it is increased with increasing $T$. 
When $T/T_{\rm IN}\gtrsim 0.9$, 
it decreases with $T$ and it almost disappears if $T>T_{\rm IN}$. 
When $T>T_{\rm IN}$, the system is in the isotropic state, 
and it does not have the long-range order. 
Thus, it is reasonable that $\Delta \mathcal{E}$ vanishes 
when $T>T_{\rm IN}$. 
When $T<T_{\rm IN}$, 
on the other hand, 
it is rather striking that the energy difference 
shows the non-monotonic dependences on $T$, 
in spite of that the long-range order and the resultant 
elasticity are reduced monotonically with increasing $T$ 
(see Fig.~\ref{fig1}). 

We plot the energy difference $\Delta \mathcal{E}$ 
as a function of $D$ in Fig.~\ref{fig3}(b), 
where the temperature is $T/T_{\rm IN}=0.89$. 
The cell thickness is changed. 
It is shown that the energy difference $\Delta \mathcal{E}$ is increased 
proportionally 
to the block size $D$ when $D$ is small. 
When the cell thickness is large, on the other hand, 
the energy difference is almost 
saturated. 
The saturated value becomes smaller when the liquid crystal is 
confined in the thicker cell.

\begin{figure}
\includegraphics[width=0.4\textwidth]{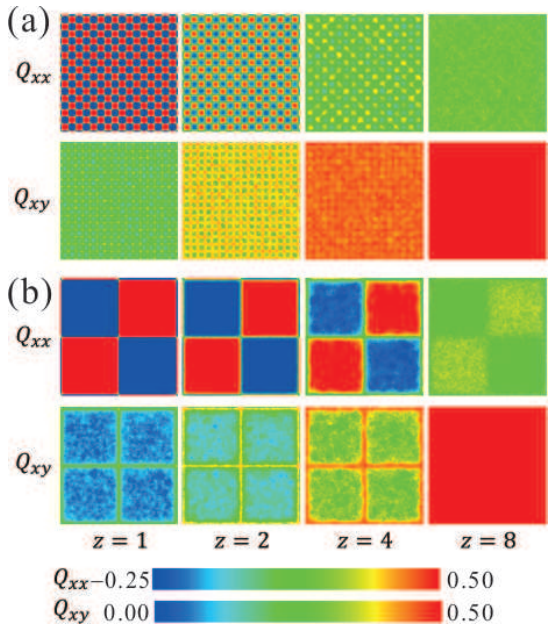}
\caption{
Snapshots of the $xx$- and $xy$ components of the tensorial 
order parameter $Q_{\mu\nu}$ in the type I cells of the 
thickness $H=8$. 
The anchoring direction on the bottom surface ($z=0$) is 
patterned like the checkerboard, while that on the top surface 
($z=9$) is homogeneously along $\hat{\bi{n}}_+$. 
The temperature is $T/T_{\rm IN}=0.89$. 
The block size is $D=8$ in (a) and $D=64$ in (b). 
Only the snapshots in a small area ($128^2$) are shown.
}
\label{fig4}
\end{figure}

In order to clarify the mechanisms of the bistable alignments, 
we calculate the spatial distribution of the nematic order parameter. 
In Fig.~\ref{fig4}, 
we show snapshots of $xx$- and $xy$ components of a 
tensorial order parameter at several planes parallel to the 
substrates. 
Using $\bi{u}_i(t')$, the tensorial order parameter 
$Q_{\mu\nu}$ is calculated  
by averaging $3/2(u_\mu u_\nu/2-\delta_{\mu\nu}/3)$ 
in  a certain period $\delta t$ as, 
\bea
Q_{i,\mu\nu}(t)=\frac{1}{\delta t}\sum_{t'=t}^{
t+\delta t-1}
\frac{3}{2}\left\{u_{i,\mu}(t')u_{i,\nu}(t')-\frac{1}{3}
\delta_{\mu\nu}\right\}, 
\ena
where $t'$ means the Monte Carlo cycle, and 
$\mu$ and $\nu$ stand for $x$, $y$ and $z$. 
In this study, we set $\delta t=10^2$, 
which is chosen so that the system is well thermalized. 
The block size is 
$D=8$ in (a) and $D=64$ in (b), 
and the cell thickness is fixed to $H=8$. 
The temperature is set to $T/T_{\rm IN}=0.89$. 
The anchoring direction at the top surface is along $\hat{\bi{n}}_+$, 
and we started the simulation with an initial condition, 
in which the director field is 
along 
$\hat{\bi{n}}_+$, 
so that the director field is likely 
to be parallel to the surface 
and along the azimuthal angle $\phi=\pi/4$ in average.  

$Q_{xx}$ near the bottom surface 
shows the checkerboard pattern 
as like as that of the imposed anchoring directions $\bi{d}_j$. 
$Q_{xy}$ inside the block domains is small 
and it is enlarged at the edges between the blocks. 
With departing from the bottom surface, 
the inhomogeneity is reduced and the director pattern 
becomes homogeneous along $\hat{\bi{n}}_+$. 
The inhomogeneities in $Q_{xx}$ and $Q_{xy}$ 
are more remarkable for the larger $D$ than 
those for the smaller $D$. 

\begin{figure}
\includegraphics[width=0.45\textwidth]{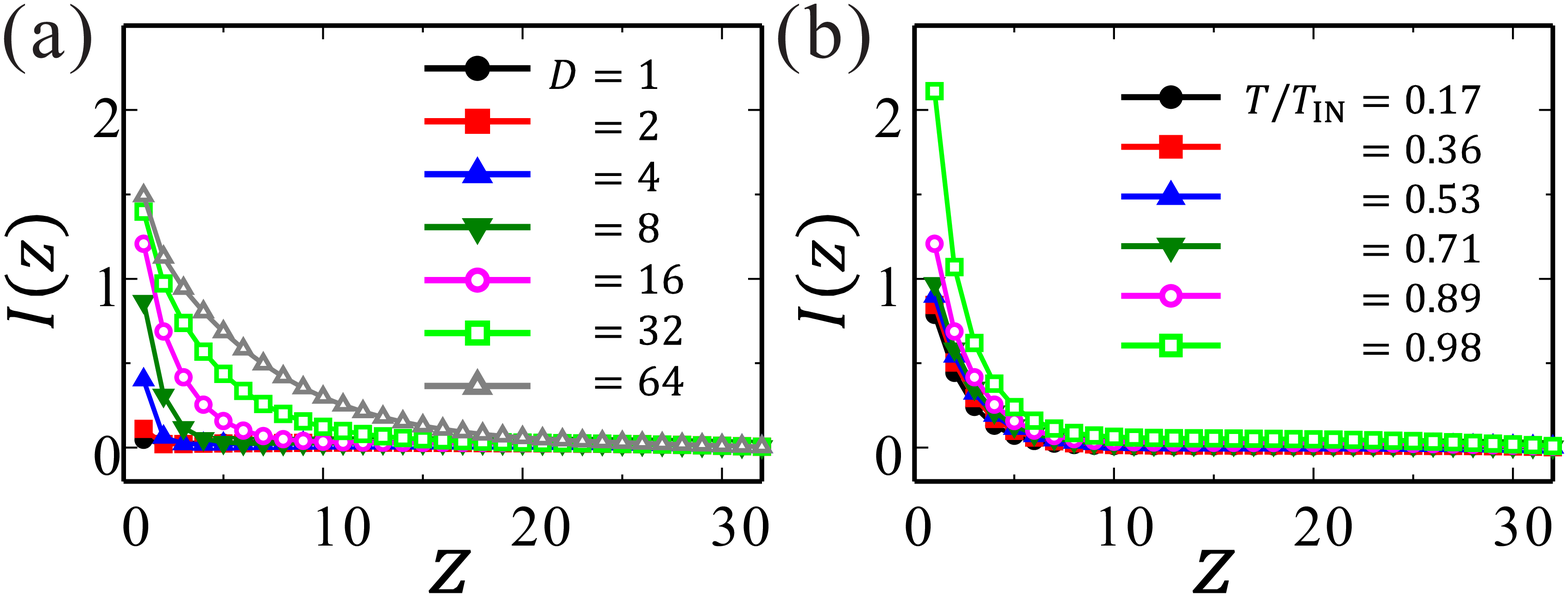}
\caption{
Profiles of the inhomogeneity of the nematic order parameter $I(z)$ along 
the cell thickness $z$. 
(a) The cell thickness is $H=32$ (type I) 
and the temperature is $T/T_{\rm IN}=0.89$. 
The block size is increased from $D=1$ to $D=64$. 
(b) The cell thickness is $H=32$ and the block size is fixed to $D=16$. 
The temperature $T$ is changed. }
\label{fig5}
\end{figure}

In Fig.~\ref{fig5}, 
we plot the corresponding 
profiles of the spatial modulations of 
the order parameter with respect to $z$. 
The degree of the inhomogeneity of $Q_{\mu\nu}$ is defined by 
\bea
I(z)&=&\frac{1}{L^2S_{\rm b}^2}
\int dx dy \{Q_{\mu\nu}(x,y,z)-\bar{Q}_{\mu\nu}(z)\}^2,
\label{eq:Iz}
\ena
where $\bar{Q}_{\mu\nu}(z)$ is the spatial average of $Q_{\mu\nu}$ 
in the $z$-plane and it is given by 
\bea 
\bar{Q}_{\mu\nu}(z)&=&L^{-2}\int dx dy Q_{\mu\nu}(x,y,z).
\label{eq:profile}
\ena
$S_{\rm b}$ is the scalar nematic parameter obtained in the bulk 
[see Fig.~\ref{fig1}(a)]. 
Since $Q_{\mu\nu}\propto S_{\rm b}$ in the bulk,
the profiles are scaled by $S^{2}_{\rm b}$ in Eq.~(\ref{eq:Iz}),
in order to see the pure degree of the inhomogeneity 
of the director field.
In Fig.~\ref{fig5}(a), we changed the block size $D$ and the 
temperature is fixed to $T/T_{\rm IN}=0.89$. 
It is shown that the degree of the inhomogeneity 
decays with $z$, and it is larger for 
larger $D$ as shown in Fig.~\ref{fig4}. 
Figure~\ref{fig5}(a) also shows that the 
decaying length is also increased with the block size $D$. 
Roughly it agrees with $D$. 
In Fig.\ref{fig5}(b), we plot the profiles of $I(z)$ 
for different temperatures with fixing $D=16$. 
It is shown that 
the spatial modulation is increased as the temperature is increased. 
This is because 
the nematic phase becomes softer as the temperature 
is increased (see Fig.~\ref{fig1}(b)). 
When the elastic modulus is small, the director field 
is distorted by the anchoring surface more largely.

Based on these numerical results, 
we consider the bistable alignments with a 
continuum elasticity theory. 
The details of the continuum theory is described in 
Appendix B. 
In our theoretical argument, 
the spatial modulation of the director field due to the 
heterogeneous anchoring 
plays a crucial role in inducing the bistable 
alignments along the diagonal directions. 
After some calculations, 
we obtained an effective anchoring energy for $D\ll H$ as 
\bea
g(\phi_0)=-\frac{cW^2D}{K}\sin^22\phi_0,
\label{eq:WDK_anc}
\ena
instead of the Rapini-Papoular anchoring energy, $-W\cos^2\phi_0/2$. 
Here $\phi_0$ is the average azimuthal angle 
of the director field on the patterned surface.
$K$ is the elastic modulus of the director field 
in the one-constant approximation of the elastic theory, 
and $W$ represents the anchoring strength in the 
continuum description. 
$c$ is a numerical factor, which is estimated as 
$c\cong 0.085$ when $H/D$ is large. 
$g(\phi_0)$ has a fourfold symmetry and is lowered 
at $\phi_0=\pm \pi/4$ and $\pm 3\pi/4$. 
The resulting energy difference per unit area is given by  
\bea
\Delta \mathcal{E}_{\rm th}=
\frac{\pi^2K}{32H\{1+K^2/(8cW^2DH)\}}.
\label{eq:E_th}
\ena 

First we discuss the dependence of the energy difference 
on the block size $D$.
Equation~(\ref{eq:E_th}) indicates that the energy difference 
behaves as $\Delta \mathcal{E}_{\rm th}\approx \pi^2 cW^2D/(4K)$, 
which is increased linearly with $D$,  
when $D$ is sufficiently small. 
If $D$ is large enough, on the other hand, 
the energy difference converges to $\Delta \mathcal{E}_{\rm th}
\approx \pi^2 K/(32H)$. 
The latter energy difference agrees with the deformation 
energy of the 
director field, which twists along the $z$ axis by $\pm \pi/4$. 
It is independent of $D$, but 
is proportional to $H^{-1}$. 
The asymptote behaviors for small and large $D$ 
are consistent with the numerical results 
shown in Fig.~\ref{fig3}(b).

Next we consider the dependence of $\Delta \mathcal{E}$ on the 
temperature. 
Equation~(\ref{eq:E_th}) also suggests 
$\Delta \mathcal{E}_{\rm th}$ is proportional to $W^2D/K$ 
when $W^2DH/K^2$ is small. 
We have speculated the anchoring strength 
is simply proportional to the nematic order as 
$W\propto S_{\rm b}$. 
If so, the energy difference is expected to be 
independent of $S_{\rm b}$ 
as $\Delta \mathcal{E}_{\rm th}\propto W^2/K 
\propto S^0_{\rm b}$, 
since $K$ is roughly proportional to $S_{\rm b}^2$. 
This expectation is inconsistent with the dependence of 
the numerical results of $\Delta \mathcal{E}$ in Fig.~\ref{fig3}(a). 
A possible candidate mechanism in explaining this discrepancy 
is that 
we should use the nematic order on the surface $S_{\rm w}$, 
instead of $S_{\rm b}$, for estimating $W$. 
Since $S_{\rm w}$ is dependent on $T$ more weakly 
than $S_{\rm b}$ near the transition temperature [see Fig.~\ref{fig1}(a)], 
$W^2/K$ can be increased with $T$. 
The curve of $S_{\rm w}^2/K$ is drawn in Fig.~\ref{fig1}(b). 
Thus, 
the director field is more largely deformed near $T_{\rm IN}$ as shown 
in Fig.~\ref{fig5}(b), 
so that the resulting energy difference shows the 
increase with $T$. 

Also, Fig.~\ref{fig3}(a) shows 
$\Delta \mathcal{E}$ turns to decrease to zero, 
when we approach to $T_{\rm IN}$ more closely. 
In the vicinity of $T_{\rm IN}$, 
$K$ is so small that $W^2/K$ becomes large. 
Then Eq.~(\ref{eq:E_th}) behaves as 
$\Delta\mathcal{E}_{\rm th} \propto K/H$. 
It is decreased to zero as $K$ with approaching to $T_{\rm IN}$. 
In Fig.~\ref{fig3}(a), we draw the theoretical 
curve of Eq.~(\ref{eq:E_th}) with taking into account 
the dependences of $W$ and $K$ on the 
temperature. 
Here we assume $W=W_0 S_{\rm w}$ with $W_0$ being a 
constant. 
The theoretical curve reproduces the non-monotonic 
behavior of the energy difference qualitatively. 
After the plateau of $\Delta \mathcal{E}_{\rm th}$ 
in the lower temperature region, 
it is increased with $T$. 
Then it turns to decrease to zero when the temperature 
is close to the transition temperature. 
Here we use $W_0=0.3$, which is chosen to adjust 
the theoretical curve to the numerical result.

\subsection{Switching dynamics}

Next we confine the nematic liquid crystals in the 
type II cells, both the surfaces of which 
are patterned as checkerboard. 
As indicated by Eq.~(\ref{eq:WDK_anc}), 
each checkerboard surface gives rise to the effective 
anchoring effect with the fourfold symmetry. 
Hence, 
the director field is expected to 
show the in-plane bistable alignments along 
$\bi{n}_+$ or $\bi{n}_-$ also in the type II cells.

\begin{figure}
\includegraphics[width=0.25\textwidth]{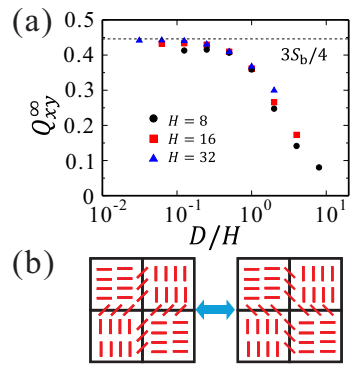}
\caption{
(a)~Dependences of the averaged order parameter on the block size. 
The cell thickness is $H=16$ (type II). 
The temperature is changed. 
(b)~Schematic pictures of the director field near the mid-plane 
in the type II cell of $D\gg H$.}
\label{fig6}
\end{figure}

Figure~\ref{fig6}(a)
plots 
the spatial average of the 
$xy$ component of $Q_{\mu\nu}$ at equilibrium 
with respect to the block size. 
The equilibrium value of $\av{Q_{xy}}$ is 
estimated as $Q_{xy}^{\infty}=\av{Q_{xy}}|_{t=5\times 10^4}$ 
in the simulations with no external field. 
As the initial condition, we employ the director field 
homogeneously aligned along $\hat{\bi{n}}_+$, 
so that $Q_{xy}^{\infty}$ is likely to be positive. 
In Fig.~\ref{fig6}(a), 
we also draw a line of $3S_{\rm b}/4$, 
which corresponds to the bulk nematic order when the 
director field is along $\bi{n}_+$. 
It is shown that 
$Q_{xy}^{\infty}$ is roughly constant and is 
close to $3S_{\rm b}/4$ for $D\ll H$. 
It is reasonable since the inhomogeneity of the director field 
is localized within $D$ from the surfaces. 
When $D>H$, on the other hand, 
$Q_{xy}^\infty$ is decreased with $D$. 
When $D\gg H$, the type II cell can be considered as a 
collection of square domains each carrying the uniform 
anchoring direction. 
Thus, the director field 
tends to be parallel to the local anchoring direction 
$\bi{d}_j$, and then, $Q_{xy}$ inside each unit block 
becomes small locally. 
Only on the edges of the block domains, the director fields 
are distorted and adopts either of the distorted 
states as schematically shown in Fig.~\ref{fig6}(b). 
With scaling $D$ by $H$, the plots of $Q_{xy}^\infty$ 
collapse onto a single curve.

\begin{figure}
\includegraphics[width=0.35\textwidth]{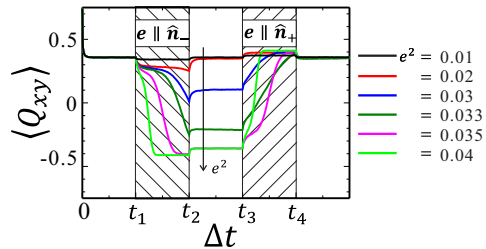}
\caption{
Time sequences of the averaged nematic order along $\hat{\bi{n}}_+$ 
in the in-plane switching processes. 
At $t=0$, 
the director field is completely aligned along 
$\bi{\bi{n}}_+$. 
In time intervals $1\times 10^4 \le t<2\times 10^4$ and 
$3\times 10^4\le t <4\times 10^4$, the external field is applied 
along $\hat{\bi{n}}_-$ and $\hat{\bi{n}}_+$, respectively. 
The strength of the external field $e$ is changed. 
The temperature is $T/T_{\rm IN}=0.89$, and the type II cell of $D=16$ 
and $H=16$ is employed. }
\label{fig7}
\end{figure}

Then we consider the switching dynamics of the director field 
between the two stable alignments with imposing in-plane 
external fields $\bi{e}$ in the type II cells. 
In Fig.~\ref{fig7}, 
we plot the spatial average of the $xy$ component of the 
order parameter $\av{Q_{xy}}$ 
in the processes 
of the director switching. 
The cell size is $H=16$, the block size is $D=16$ and 
the temperature is $T/T_{\rm IN}=0.89$. 
At $t=0$, we start the Monte Carlo simulation with the same initial condition, 
in which the director field is homogeneously aligned along 
$\hat{\bi{n}}_+$, 
in the absence of the external field. 
As shown in Fig.~\ref{fig7}, 
the nematic order is relaxed to 
a certain positive value, 
which agrees with $Q_{xy}^{\infty}$ in Fig.~\ref{fig6}(a). 
From $t_1=10^4$, 
we then impose an in-plane external field along 
$\hat{\bi{n}}_-$, and turn it off at $t_2=2\times 10^4$. 
After the system is thermalized during $t=2\times 10^4$ and 
$t_3=3\times 10^4$ with no external field, 
we apply the second external field along $\hat{\bi{n}}_+$ 
from $t_3=3\times 10^4$ 
until $t_4=4\times 10^4$. 
We change the strength of the external field $e$.

When the external field is weak ($e^2\le 0.03$), 
the averaged orientational order is slightly reduced by the external field, 
but it recovers the original state after the field is removed. 
After a strong field ($e^2\ge 0.04$) is applied and 
is removed off, on the other hand, 
$\av{Q_{xy}}$ 
is relaxed to another steady state value, which is close to 
$-Q_{xy}^\infty$. 
This new state of the negative $\av{Q_{xy}}$ 
corresponds to the other bistable alignment along 
$\hat{\bi{n}}_-$.
After the second field along 
$\hat{\bi{n}}_+$ is applied, 
the averaged orientational order $\av{Q_{xy}}$ 
comes back to the positive original value, $+Q_{xy}^\infty$.

\begin{figure}
\includegraphics[width=0.5\textwidth]{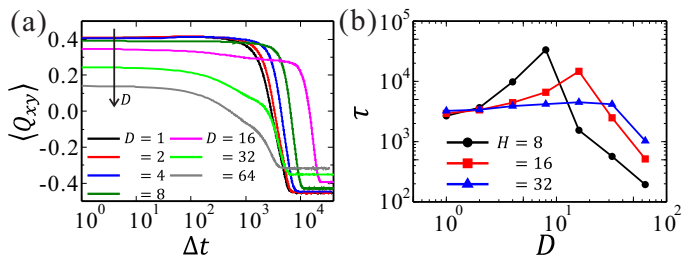}
\caption{
(a) 
Time sequences of the averaged nematic order parameter $\av{Q_{xy}}$ 
in the switching process. 
Before $\Delta t=0$, the director field is aligned 
along $\hat{\bi{n}}_+$ in average. 
After $\Delta t=0$, the in-plane external field is applied along 
$\hat{\bi{n}}_-$. 
The temperature is $T/T_{\rm IN}=0.89$ and the cell thickness is $H=16$ 
(type II). 
The strength of the external field $e$ is changed. 
(b) Plots of the characteristic time $\tau$ of the switching process 
with respect to the block size $D$. 
$\tau$ is defined as $\av{Q_{xy}(\Delta t=\tau)}=0$. 
The temperature is $T/T_{\rm IN}=0.89$ and the thickness of 
type II cell is changed. }
\label{fig8}
\end{figure}

In Fig.~\ref{fig8}(a), 
we show the detailed relaxation behaviors of 
$\av{Q_{xy}}$ in the first switching after $t_1$. 
$\Delta t$ means the elapsed time in the first switching, 
that is $\Delta t=t-t_1$. 
Here 
we change the block size $D$, while 
we fix the external field at $e^2=0.03$ and 
the cell thickness $H=16$ (type II). 
We note that $\av{Q_{xy}}$ at $\Delta t=0$ 
depends on $D$ as indicated in Fig.~\ref{fig6}(a). 
In Fig.~\ref{fig8}(a), it is shown that the switching rate depends 
also on the block size $D$. 
Notably, the dependence of the switching behavior is not 
monotonic against $D$.

In Fig.~\ref{fig8}(b), we plot the characteristic switching time 
$\tau$ with respect to the block size $D$ in the cells of 
$H=8$, $16$ and $32$. 
The temperature and the field strength are the same those for 
Fig.~\ref{fig8}(a).
The characteristic time $\tau$ is defined such that the average orientational 
order is equal to zero at $\tau$, $\av{Q_{xy}}(\Delta t=\tau)=0$. 
Figure~\ref{fig8}(b) shows the characteristic time is maximized 
when the 
block size is comparable to the cell thickness. 
When $D<H$, the switching process is slowed down as the block size 
is increased. 
On the other hand, it is speeded up with $D$ when $D>H$. 
In Fig.~\ref{fig8}(b), it is suggested that 
the dependence of $\tau$ on $D$ becomes less significant 
as $H$ is increased.

\begin{figure}
\includegraphics[width=0.45\textwidth]{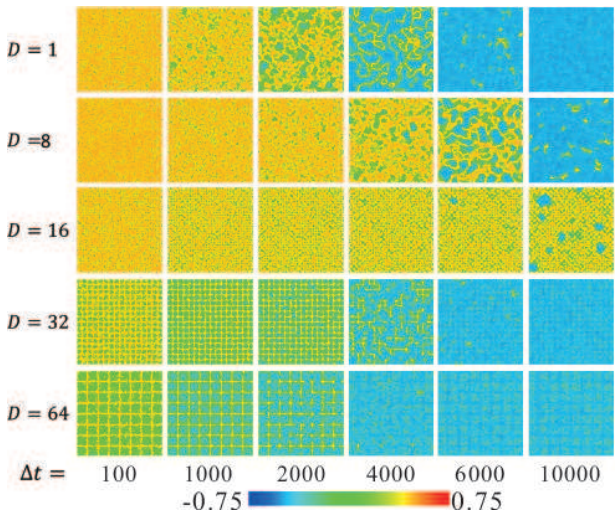}
\caption{
Snapshots of the nematic order parameter $Q_{xy}(x,y)$ at $z=8$ 
in the type II cell of $H=16$. 
The director field in yellow regions is along $\hat{\bi{n}}_+$ and 
that in blue regions is along $\hat{\bi{n}}_-$.  
The temperature is $T/T_{\rm IN}=0.89$ and the block size $D$ 
is changed. 
}
\label{fig9}
\end{figure}

Figure~\ref{fig9} depicts 
snapshots of 
$Q_{xy}(t)$
at the midplane ($z=H/2$) during the first switching process. 
The parameters are the same as those in Fig.~\ref{fig8}(a), 
so that the pattern evolutions 
correspond to the curves of $\av{Q_{xy}}$ in Fig.~\ref{fig8}(a). 
Figure~\ref{fig9} shows that the switching behavior  
is slowed down when $D$ is comparable to $H$, 
in accordance with Fig.~\ref{fig8}(b). 
When $D<H$, the snapshots implies the switching proceeds 
via nucleation and growth mechanism. 
From the sea of the positive $Q_{xy}$, where the director is 
aligned along $\hat{\bi{n}}_+$, 
the droplets of the negative $Q_{xy}$ are nucleated. 
They grow with 
time and cover the whole area eventually. 
Under the external field along $\hat{\bi{n}}_-$, 
the alignment of the director field along $\hat{\bi{n}}_-$ 
is more preferred than that along $\hat{\bi{n}}_+$. 
Because of the energy barrier between these bistable alignments, 
the director field cannot change its 
orientation to $\hat{\bi{n}}_-$ smoothly under 
a weak external field. 
From Eq.~(\ref{eq:WDK_anc}), the energy barrier 
for the local swiching of the director field 
between the two stable states is given by 
$\Delta \mathcal{F}=8D^2 \{g(\phi_0=0)-g(\pi/4)\}=8cW^2D^3/K$, 
when $D<H$. 
Thus, the slowing down of the switching process 
with $D$ is considered to be 
attributed to the enhancement of the energy barrier. 
Here we note that 
a critical field strength 
for the thermally activated switching cannot be defined unambiguously.
Since the new alignment is energetically preferred over the 
original one even under a weak field, 
the director configuration will change its orientation 
if the system is annealed for a sufficiently long period. 
When the field strength is moderate ($e^2\cong 0.035$), 
the averaged order 
goes to an intermediate value, 
neither of $Q_{xy}^\infty$ or $-Q_{xy}^\infty$ in Fig.~\ref{fig7}. 
Such intermediate values of $\av{Q_{xy}}$ reflect large scale 
inhomogeneities of the bistable alignments (see Fig.~\ref{fig9}). 
At each block, the director field adopts either of the 
two stable orientations. 
The pattern of the intermediate $\av{Q_{xy}}$ 
depends not only on the field strength, but also 
on the annealed time. 
Under large external fields, 
on the other hand, 
the energy barrier between the two 
states can be easily overcome, so that  
the switching 
occurs without arrested at the initial orientation 
(not shown here). 

Regarding the local director field, which adopts 
either of the two stable orientations ($\hat{\bi{n}}_+$ and 
$\hat{\bi{n}}_-$), as a binarized spin 
at the corresponding block unit, we found a similarity 
of the domain growth in our system and that in a two-dimensional 
Ising model subject to an external magnetic field. 
If the switching of the director field occurs locally 
only at each block unit, 
there is no correlations between the director fields 
in the adjacent block units. 
Therefore, the nucleation and growth switching behavior 
implies the director field at a block unit prefers to be aligned 
along the same orientation as those at the adjacent block units. 

We observed  string-like patterns 
as shown at $\Delta t=4000$ for $D=1$ in 
Fig.~\ref{fig9}. 
Here we note that they are not 
disclinations of the director field. 
They represent domain walls perpendicular to the substrates. 
In the type II cells, we have not observed 
any topological defects, 
although topological defects are sometimes stabilized 
in the frustrated cell \cite{Araki_NM_2011}. 
The string-like patterns remain rather stable transiently. 
On the other hand, 
such string-like patterns are not observed in the switching process 
in the Ising model. 
This indicates  the binarized spin description of the bistable director 
alignments may be not adequate. 
Under the external field along $\hat{\bi{n}}_-$, 
the director field rotates to the new orientation 
clockwise or counter-clockwise. 
New domains, which appear via the clockwise rotations, 
have some mismatches against those through the counter-clockwise 
rotations. 
The resulting boundaries between the incommensurate domains 
are formed and 
tend to suppress the coagulations of them more and less, 
although the corresponding energy barriers are not so large.

When $D>H$, 
the switching occurs in a different way. 
The director rotations are localized around the 
edges of the blocks 
as indicated in Fig.~\ref{fig6}(b). 
As $D$ is increased, the amount of the 
director field that reacts to the field 
is reduced. 
Although 
the director fields 
around the centers of the blocks
do not show any switching behaviors 
before and after the field application, 
they are distorted to orient slightly 
toward the field. 
It is considered that the distortion of the 
director field inside the blocks effectively reduces 
the energy barrier against the external field.
We have not succeeded in explaining 
the mechanism of the reduction of the 
switching time with $D$. 
When $D>H$, the inhomogeneous director field 
contains higher Fourier modes of the distortion. 
The energy barrier for 
each Fourier mode becomes lower for the higher 
Fourier modes [see Eq.~(\ref{eq:EE}). 
Thus, such higher Fourier modes are more active 
against the external field and they would 
behave as a trigger of the switching process.

\section{Conclusion}

In this article, we studied nematic liquid crystals confined 
by two parallel checkerboard substrates by means of 
Monte Carlo simulation of the Lebwohl-Lasher model. 
As observed experimentally by Kim {\it et al.},
we found the director field in the bulk shows the bistable alignments, 
which are along 
either of the two diagonal axes. 
We attribute the bistability of the alignments to the 
spatial modulation of the director field near the substrates. 
Based on the elastic theory, we derived 
an effective 
anchoring energy with the fourfold symmetry (Eq.~(\ref{eq:WDK_anc})). 
Its anchoring strength is expected to behave as 
$W^2D/K$, when the block size $D$ is smaller than 
the cell thickness. 
As the temperature is increased to the isotropic-nematic transition 
temperature, the elastic modulus $K$ of the nematic phase is reduced 
so that the director field is deformed near the substrates more largely. 
With this effective anchoring effect, we can explain 
the non-monotonic dependence of the energy stored in this cell 
qualitatively.

We also studied the switching dynamics of the director configuration 
with imposing in-plane external fields. 
Usually, the switching is considered to be associated with the 
actual breaking of the anchoring condition. 
Thus, the energy barrier for the switching is expected to be 
proportional to $W$ \cite{Kim_APL_2001}. 
In this article, we propose another possible mechanism of the 
switching, in which the anchoring condition is not necessarily broken. 
Since the energy barrier is increased with the block size, 
the switching dynamics notably becomes slower when the block size is 
comparable to the cell thickness. 

By solving $\Delta \mathcal{F}=\Delta \epsilon E^2/2\times (4D^2H)$, 
we obtain a characteristic strength of the electric field $E$ 
as 
$E_{\rm c}\cong \{8cW^2D/(\Delta \epsilon KH)\}^{1/2}$, 
where $\Delta \epsilon$ is the anisotropy of the dielectric 
constant [see Eq.~(\ref{eq:Frank})]. 
If we apply an in-plane external field larger than $E_{\rm c}$, 
the switching occurs rather homogeneously 
without showing the nucleation and growth processes. 
This characteristic strength is decreased with decreasing $D$, 
so that the checkerboard pattern of smaller $D$ is preferred 
to reduce the field strength. 
With smaller $D$, however, the stability of the two preferred 
orientations is reduced. 
If the effective anchoring energy is lower than the thermal energy, 
the 
bistable alignment will be destroyed by the thermal fluctuation. 
In this sense, the block size $D$ should be larger 
than $D_{\rm c}\approx (Kk_{\rm B}T/8cW^2)^{1/3}$, where we 
assumed that the switching occurs locally in each block, 
that is $\Delta \mathcal{F}\approx k_{\rm B}T$.
For a typical nematic liquid crystal with $K=1\,{\rm pN}$ and 
$W=10^{-5} \,{\rm J/m^2}$ at room temperature $T=300\,{\rm K}$, 
it is estimated as $D_{\rm c}\cong 34\,{\rm nm}$.

In our theoretical argument, 
we assumed the one-constant approximation of the elastic modulus. 
However, the director field cannot be described 
by a single deformation mode in the above cells. 
The in-plane splay and bend deformations are localized 
within the layer of $D$ near the surface. 
On the other hand, 
the twist deformation is induced by the external field 
along the cell thickness direction. 
If the elastic moduli for the three deformation modes 
are largely different from each other, 
our theoretical argument would be invalid. 
We need to improve both the theoretical and numerical 
schemes to consider such dependences more correctly. 
Also, we considered only the checkerboard substrates. 
But, it is interesting and important 
to design other types of patterned surfaces \cite{Kim_Nature_2002} 
to append more preferred functions, 
such as faster responses against the external field, 
to liquid crystal devices. 
We hope to report a series of such studies in the near future.

\section*{acknowledgements}

We acknowledge valuable discussions with J. Yamamoto, H. Kikuchi, 
I. Nishiyama, K. Minoura and T. Shimada. 
This work was supported by KAKENHI 
(No. 25000002 and No. 24540433). Computation was done using the
facilities of the Supercomputer Center, the Institute for Solid
State Physics, the University of Tokyo.

\appendix

\section{Estimations of the nematic order and the elastic moduls}

In this appendix, 
we estimate the scalar nematic order parameter 
and the elastic modulus in Fig.~\ref{fig1} 
from the Monte Carlo simulations with Eq.~(\ref{eq:Hamiltonian}) 
\cite{Lebwohl_PRA_1972,Chiccoli_PhysicaA_1988,Cleaver_PRA_1991,Mondal_PLA_2003,
Marguta_PRE_2011,Maritan_PRL_1994,Priezjev_PRE_2001,Araki_NM_2011}. 
First we consider the bulk behaviors of nematic liquid crystals, 
which are described by the Lebwohl-Lasher potential. 
Here we remove the surface sites $\mathcal{S}$ and employ 
the periodic boundary conditions for all the axes ($x,y$, and $z$). 
We use the initial condition along $\bi{u}_i=(1,0,0)$ and thermalize the 
system with the heat bath sampling. 
The simulation box size is $L^3$ with $L=128$. 

It is well known that this Lebwohl-Lasher spin model describes 
the first-order transition between isotropic and nematic phases. 
In Fig.~\ref{fig1}, we plot the $xx$ component of the tensorial order 
parameter after the thermalization ($t\le 5\times 10^4$) 
as a function of $T$. 
Since the initial condition is along the $x$ axis, 
the director field is likely to be aligned along the $x$ axis.
We here regard $\av{Q_{xx}}$ 
as the scalar nematic order parameter $S_{\rm b}$.
We see an abrupt change of $S_{\rm b}$ around $k_{\rm B}T_{\rm IN}\approx 
1.12\epsilon$, which is consistent 
with previous studies \cite{Priezjev_PRE_2001}. 
Above $T_{\rm IN}$, the nematic order almost vanishes, while 
it is increased with decreasing $T$ when $T<T_{\rm IN}$. 

In the nematic phase ($T<T_{\rm IN}$), 
the director field $\bi{n}$ can be defined. 
Because of the thermal noise, 
the local director field is fluctuating around the 
average director field, reflecting the elastic modulus. 
The elastic modulus of the director field is obtained 
by calculating the scattering function of the 
tensorial order parameter as \cite{Cleaver_PRA_1991},
\bea
\av{|\tilde{Q}_{x\mu}(\bi{q})|^2}_T=
\frac{k_{\rm B}T}{A+4L_Q\sin^2 |\bi{q}|a/2},
\ena
for $\mu=y$ and $z$. 
$\tilde{Q}_{x\mu}(\bi{q})$ is the Fourier component 
of $Q_{x\mu}$ at a wave vector $\bi{q}$. 
$a$ is the lattice constant, and $\av{\cdots}_T$ means 
the thermal average. 
$A$ and $L_Q$ are the coefficients appearing in the 
free energy functional for $Q_{\mu\nu}$. 

In the case of $T<T_{\rm IN}$, the scattering function goes to 
zero for $|\bi{q}|a\cong 0$. 
Then, we obtain the coefficient $L_Q$ by fitting 
$\av{|\tilde{Q}_{x\mu}|^2}^{-1}_T$ 
with $4(k_{\rm B}T)^{-1}L_Q\sin^2 |\bi{q}|a/2$. 
$L_Q$ is proportional to the elastic modulus $K$ of the director field 
$\bi{n}$ as $L_Q=KS^2_{\rm b}$. 
In Fig.~\ref{fig1}(b), 
the elastic modulus $K$ is plotted with respect to $T$. 
It is decreased with increasing $T$, if $T>T_{\rm IN}$. 
This indicates the softening of the nematic phase near the transition 
temperature. 

Next we consider the effect of the surface term. 
The surface effect not only induces the angle dependence 
of the anchoring effect in the nematic phase, but also leads to the wetting 
effect of the nematic phase to the surface in the isotropic phase 
\cite{Sheng_PRA_1982}. 
We set homogeneous surfaces of $w=\epsilon$ 
at $z=0$ and $z=H+1$ as in the main text. 
The anchoring direction is $\bi{d}_j=(1,0,0)$. 
The periodic boundary conditions are imposed for the $x$- and $y$-directions
and the initial condition is along $\bi{u}_i=(1,0,0)$. 
The profile of $\bar{Q}_{xx}$ (not shown here) indicates 
$\bar{Q}_{xx}$ at $z=1$ becomes larger than that in the bulk, $S_{\rm b}$. 
This value is the surface order $S_{\rm w}$, which is also plotted in 
Fig.~\ref{fig1}(a) with red open squaress. 
Notably, $S_{\rm w}$ remains a finite value 
even when $T>T_{\rm IN}$. 
In Fig.~\ref{fig1}(a), we cannot see any drastic change of $S_{\rm w}$, 
which is continuously decreased with $T$.

\section{Analysis with Frank elasticity theory}

Here, we consider the nematic liquid crystal 
confined in the checkerboard substrate 
on the basis of the Frank elasticity theory. 
The checkerboard substrate is placed at $z=0$, 
while we fix the director field at the top surface 
like the type I cells employed in the simulations. 
The free energy of the nematic liquid crystal 
is given by 
\bea
\mathcal{F}&=&\frac{K}{2}\int d\bi{r}
(\n \bi{n})^2-\frac{\Delta \epsilon}{2}\int d\bi{r}(\bi{n}\cdot\bi{E})^2
\nonumber\\
&&-W\int_{z=0}dxdy(\bi{n}\cdot \bi{d})^2,
\label{eq:Frank}
\ena
where $\bi{n}$ is the director field. 
The first term in the right hand side of Eq.~(\ref{eq:Frank}) 
is the elastic energy. 
Here we employ the one-constant approximation 
with the elastic modulus $K$. 
$\bi{E}$ and $\Delta\epsilon$ are external electric field 
and the anisotropy of the dielectric constant. 
Here
we do not consider the effect of the electric field. 
The third term in Eq.~(\ref{eq:Frank}) 
represents the anchoring energy in 
the Rapini-Papoular form. 
$W$ is the anchoring strength and $\bi{d}$ 
is the preferred direction on the surface at $z=0$. 
For the checkerboard substrates, 
we set $\bi{d}$ according to Eq.~(\ref{eq:block}). 

At the top surface, we fix the director field 
as $\bi{n}(z=H)=\bi{d}_{\rm t}(\cos \phi_{\rm t},\sin \phi_{\rm t},0)$, 
and the bottom surface also prefers the planar anchoring. 
From the symmetry, therefore, 
we assume that the director field in the bulk lies parallel 
to the substrates everywhere. 
Then, we can write it only with the azimuthal angle $\phi$ as 
\bea
\bi{n}=(\cos \phi,\sin \phi,0). 
\ena
Also, we assume that the director field 
is periodic for $x$ and $y$ directions, 
so that we only have to consider the 
free energy in the unit block $(0\le x,\, y \le 2D)$. 
With these assumptions, the free energy per unit area
is written as 
\bea
&&\mathcal{E}
=\frac{K}{8D^2}\int_0^{2D}dx\int_0^{2D}dy\int_0^Hdz (\n\phi)^2
\nonumber\\
&&
-\frac{W}{2D^2}
\left.\int_0^Ddx
\left\{
\int_0^Ddy \sin^2\phi
+\int_D^{2D}dy\cos^2\phi
\right\}\right|_{z=0}.
\nonumber\\
\label{eq:Frank2}
\ena
In the equilibrium state, 
the free energy is minimized with respect to $\phi(x,y,z)$. 
Inside the cell ($0<z<H$), 
the functional derivative of $\mathcal{E}$ gives 
the Laplace equation of $\phi$ as
\bea
\frac{\delta \mathcal{E}}{\delta \phi}=-K\n^2\phi=0. 
\label{eq:Laplace}
\ena
From the symmetry argument, 
we have its solution as 
\bea
\phi(x,y,z)&=&\phi_0+(\phi_{\rm t}-\phi_0)z/H+
\Delta(x,y,z),
\\
\Delta(x,y,z)&=&
\sum_{m,n=0}^\infty\Delta_{mn}
\sin\frac{(2m+1)\pi x}{D}\sin \frac{(2n+1)\pi y}{D}
\nonumber\\
&&\times 
\sinh(\pi \gamma_{mn}(H-z)/D), \nonumber\\
\gamma_{mn}&=&\sqrt{(2m+1)^2+(2n+1)^2}
\label{eq:Delta}
\ena
where 
$\phi_0$ and $\Delta_{mn}$ are determined later. 

It is not easy to calculate the second term 
in Eq.~(\ref{eq:Frank2}) analytically. 
Assuming $|\Delta|\ll 1$, we approximate $\sin^2\phi$ as 
\bea
\sin^2(\phi_0+\Delta) 
\approx \sin^2\phi_0+\Delta \sin 2\phi_0 +\Delta^2\cos 2\phi_0,
\label{eq:sin}
\ena
Then, we obtain the free energy per unit area as 
\bea
\mathcal{E}
&=&
\frac{K}{2H}(\phi_{\rm t}-\phi_0)^2-\frac{W}{2}\nonumber\\
&&+\sum_{m,n}
\left[
\frac{\pi K \Delta_{mn}^2\sinh(2\pi \gamma_{mn}H/D)}{16D}
\right.\nonumber\\
&&\left.
-\frac{4W\Delta_{mn}\sin2\phi_0 \sinh(\pi \gamma_{mn}H/D)}{(2m+1)(2n+1)\pi^2}
\right]. 
\label{eq:EE}
\ena

First we minimize the free energy with respect to $\Delta_{mn}$ by solving  
$\partial \mathcal{E}/\partial \Delta_{mn}=0$. 
Then, we have 
\bea
\Delta_{mn}=\frac{16WD\mathrm{sech}
(\pi \gamma_{mn}H/D)\sin2\phi_0}{(2m+1)(2n+1)\gamma_{mn}\pi^3 K},
\ena
and
\bea
\mathcal{E}&\approx&\frac{K}{2H}(\phi_{\rm t}-\phi_0)^2-\frac{W}{2}
-c\frac{W^2D}{K}\sin^22\phi_0,
\label{eq:E}\\
c&=&
\sum_{mn}\frac{32\tanh(\pi \gamma_{mn}H/D)}{(2m+1)^2(2n+1)^2\pi^5\gamma_{mn}}.
\ena
In the limit of $H\gg D$, $c$ converges to $c\approx 0.085$, 
while it behaves as $c\approx 0.5H/D$ if $H\ll D$. 
Since $c$ is positive, 
the last term in the right-hand side of Eq.~(\ref{eq:E}) 
represents an effective anchoring condition  
[Eq.~(\ref{eq:WDK_anc})] in the main text. 
It indicates 
the director field tends to be along the diagonal axes of 
the checkerboard surface, $\phi_0=\pm \pi/4$. 
Then, we minimize $\mathcal{E}$ with respect to $\phi_0$ 
and obtain 
\bea
\mathcal{E}=-\frac{W}{2}-\frac{cW^2D}{K}+\frac{K}{2H}\frac{(\phi_{\rm t}\mp\pi/4)^2}{1+K^2/(8cW^2DH)}.
\label{eq:EEE}
\ena
It corresponds to the plots in Fig.~\ref{fig2}(a). 
Here we assumed $|\phi_0\mp \pi/4|\ll 1$, 
so that $\sin^22\phi_0\cong 1-4(\phi_0\pm \pi/4)^2$. 
The resulting energy difference is obtained as
\bea
\Delta \mathcal{E}=\frac{\pi^2K}{32H\{1+K^2/(8cW^2DH)\}}.
\label{eq:DeltaE}
\ena

In the strong anchoring limit, 
we can obtain $\phi$ rigorously as 
\bea
&&\phi(x,y,z)=(\phi_{\rm t}\mp\pi/4)z/H \pm\pi/4\nonumber\\
&&+\frac{4}{\pi}\sum_{m,n} \frac{1}{(2m+1)(2n+1)}
\sin\frac{(2m+1)\pi x}{D}\sin \frac{(2n+1)\pi y}{D}\nonumber\\
&&\times \sinh(\pi \gamma_{mn}(H-z)/D)
\ena
Its energy difference is then given by $\Delta \mathcal{E}=\pi^2K/(32H)$. 
It is consistent with Eq.~(\ref{eq:DeltaE}) 
in the limit of $W\rightarrow \infty$.

\bibliographystyle{apsrev4-1}
\bibliography{nematic_confine}

\end{document}